\documentclass{article}
\usepackage[margin=1.25in]{geometry}
\usepackage{graphicx} 
\usepackage{graphics}
\usepackage{amsmath}
\usepackage{amssymb}
\usepackage{bbm}
\usepackage{mathtools}
\usepackage{amsfonts}
\usepackage{bbold}
\usepackage{stmaryrd}
\usepackage{relsize}
\usepackage[shortlabels]{enumitem}
\usepackage{hyperref}
\usepackage[dvipsnames]{xcolor}
\usepackage{verbatim}
\usepackage{caption, subcaption}
\usepackage{float}
\usepackage{overpic}
\usepackage{adjustbox}
\usepackage{titlesec}
\titleformat*{\section}{\large\bfseries}
\titleformat*{\subsection}{\normalsize\bfseries}
\titleformat*{\subsubsection}{\normalsize\bfseries}
\titleformat*{\paragraph}{\normalsize\bfseries}
\titleformat*{\subparagraph}{\large\bfseries}
\usepackage{tikz-cd,mathtools}
\tikzset{mydescription/.style={anchor=center,fill=white}}
\usetikzlibrary{bbox}

\usepackage{tikz-cd} 
\usetikzlibrary{calc}
\tikzset{curve/.style={settings={#1},to path={(\tikztostart)
    .. controls ($(\tikztostart)!\pv{pos}!(\tikztotarget)!\pv{height}!270:(\tikztotarget)$)
    and ($(\tikztostart)!1-\pv{pos}!(\tikztotarget)!\pv{height}!270:(\tikztotarget)$)
    .. (\tikztotarget)\tikztonodes}},
    settings/.code={\tikzset{quiver/.cd,#1}
        \def\pv##1{\pgfkeysvalueof{/tikz/quiver/##1}}},
    quiver/.cd,pos/.initial=0.35,height/.initial=0}

\tikzcdset{%
    triple line/.code={\tikzset{%
        double distance = 3pt, 
        double=\pgfkeysvalueof{/tikz/commutative diagrams/background color}}},
    quadruple line/.code={\tikzset{%
        double distance = 5.3pt, 
        double=\pgfkeysvalueof{/tikz/commutative diagrams/background color}}},
    Rrightarrow/.code={\tikzcdset{triple line}\pgfsetarrows{tikzcd implies cap-tikzcd implies}},
    RRightarrow/.code={\tikzcdset{quadruple line}\pgfsetarrows{tikzcd implies cap-tikzcd implies}}
}

\usepackage{amsthm}

\newtheorem{theorem}{Theorem}[section]

\newtheorem*{thm*}{Theorem}
\newtheorem{proposition}[theorem]{Proposition}
\newtheorem*{prop*}{Proposition}
\newtheorem{corollary}[theorem]{Corollary}
\newtheorem*{cor*}{Corollary}
\newtheorem{lemma}[theorem]{Lemma}
\newtheorem*{lemma*}{Lemma}

\theoremstyle{definition}
\newtheorem{definition}[theorem]{Definition}
\newtheorem*{defn*}{Definition}
\newtheorem{example}[theorem]{Example}

\newtheorem{remark}[theorem]{Remark}

\newtheorem*{rem*}{Remark}

\newtheorem{question}[theorem]{Question}
\newtheorem{hypothesis}[theorem]{Hypothesis}

\newcommand{\sVect}{\mathbf{sVect}}

\newcommand{\sAlg}{\mathbf{sAlg}}

\newcommand{\Bord}{\mathbf{Bord}}

\newcommand{\Vect}{\mathbf{Vect}}
\newcommand{\Mod}{\mathbf{Mod}}

\newcommand{\id}{\operatorname{id}}
\newcommand{\Hom}{\operatorname{Hom}}

\newcommand{\End}{\operatorname{End}}

\newcommand{\Aut}{\operatorname{Aut}}

\newcommand{\Tr}{\operatorname{Tr}}

\newcommand{\pt}{\operatorname{pt}}

\newcommand{\Ext}{\operatorname{Ext}}

\newcommand{\Fun}{\operatorname{Fun}}

\newcommand{\Pic}{\operatorname{Pic}}

\newcommand{\Alg}{\mathbf{Alg}}
\newcommand{\sW}{s\mathcal{W}}

\newcommand{\Sp}{\mathbf{Sp}}
\newcommand{\BrFus}{\mathbf{BrFus}}
\newcommand{\BrTens}{\mathbf{BrTens}}
\newcommand{\Fus}{\mathbf{Fus}}
\newcommand{\Tens}{\mathbf{Tens}}

\newcommand{\Rex}{\mathbf{Rex}}
\renewcommand{\Pr}{\mathbf{Pr}}

\newcommand{\TQFT}{\operatorname{TQFT}}
\newcommand{\ITQFT}{\operatorname{ITQFT}}

\newcommand{\Z}{\mathbb{Z}}
\newcommand{\R}{\mathbb{R}}
\newcommand{\C}{\mathbb{C}}

\newcommand{\Q}{\mathbb{Q}}

\newcommand{\tors}{\operatorname{tors}}

\definecolor{Blue} {rgb} {0.282352,0.239215,0.803921}
\definecolor{Green} {rgb} {0.133333,0.545098,0.133333}
\definecolor{Red}   {rgb} {0.803921,0.000000,0.000000}
\definecolor{Violet}{rgb} {0.580392,0.000000,0.827450}
\definecolor{darkspringgreen}{rgb}{0.09, 0.45, 0.27}

\newcounter{jfc}

  \newcounter{jfcs}

\title{Is Crane--Yetter fully extended?}
\author{Luuk Stehouwer}
\date{\today}

\begin{document}

\maketitle

\begin{abstract}
    We revisit the question of whether the Crane--Yetter topological quantum field theory (TQFT) associated to a modular tensor category admits a fully extended refinement. More specifically, we use tools from stable homotopy theory to classify extensions of invertible four-dimensional TQFTs to theories valued in symmetric monoidal 4-categories whose Picard spectrum has nontrivial homotopy only in degrees 0 and 4. We show that such extensions are classified by two pieces of data: an equivalence class of an invertible object in the target and a sixth root of unity.  Applying this result to the 4-category $\mathbf{BrFus}$ of braided fusion categories, we find that there are infinitely many equivalence classes of fully extended invertible TQFTs reproducing the Crane--Yetter partition function on top-dimensional manifolds, parametrized by a $\mathbb{Z}/6$-extension of the Witt group of nondegenerate braided fusion categories.
    This analysis clarifies common claims in the literature and raises the question of how to naturally pick out the $SO(4)$-fixed point data on the framed TQFT which assigns the input braided fusion category to the point so that it selects the Crane--Yetter state-sum.
\end{abstract}
\section{Introduction}

The purpose of this note is to highlight some subtleties regarding the question of whether certain invertible topological field theories are fully extended.
Namely, we emphasize that what it means to extend a given nonextended TQFT down to points both highly depends on the target higher category, as well as on the geometric structure on spacetime.
Some of our results are known to experts---see in particular \cite{danlens} and \cite{schommer2018tori}---but have not diffused completely into the tensor categories community to provide a practical understanding.
We aim to change this by proving potentially confusing concrete statements such as

\begin{theorem}
\label{mainth}
    Let $\BrFus$ be the $4$-category of braided fusion categories over $\C$.
    Given a modular fusion category $\mathcal{C}$ and nonzero numbers $a,b \in \C^\times$, there exist $6$ equivalence classes of four-dimensional TQFTs with target $\BrFus$ which assign $\mathcal{C}$ to a point and have partition function
    \begin{equation}
    \label{eq:arbitrarypartfn}
    Z(X^4) = a^{\chi(X)} b^{\sigma(X)},
    \end{equation}
    where $\chi$ is the Euler characteristic, $\sigma$ is the signature, and $X$ is a closed $4$-manifold.
\end{theorem}

It is well-known that \emph{invertible} nonextended four-dimensional TQFTs $\Bord_{4,3} \to \Vect$ are classified by their partition function, which is always of the form \eqref{eq:arbitrarypartfn}. 
The above theorem therefore proves in particular that any such TQFT can be extended down to points with target $\BrFus$.
However, it says much more: we can assign \emph{any} braided fusion category to the point.

This discussion shakes up some lore about how exactly the Crane--Yetter TQFT is fully extended.
More specifically, the Crane--Yetter state-sum model for a given modular fusion category $\mathcal{C}$ has as its partition function
\begin{equation}
\label{eq:partitionfunction}
Z(X) =  (\dim \mathcal{C})^{\chi(X)/2} e^{2\pi i c\sigma(X)/8}
\end{equation}
for a closed $4$-manifold $X$.
Here $\dim \mathcal{C}$ is the global dimension of $\mathcal{C}$ and $c$ its central charge.
It is separately known that every nondegenerate braided fusion category $\mathcal{C}$ defines an invertible object in $\BrFus$, and so is in particular fully dualizable.
Hence by the cobordism hypothesis there is a unique fully extended framed four-dimensional TQFT $Z$ such that $Z(*) = \mathcal{C}$.
Since $SO(4)$-fixed point data has been conjectured to be related to the ribbon structure on $\mathcal{C}$, it is very tempting to conjecture that we can somehow canonically extend this framed TQFT to a fully extended oriented theory with partition function \eqref{eq:partitionfunction}.
However, the above theorem implies not only that we can define $SO(4)$-fixed point data on this framed theory so that its partition function becomes \eqref{eq:partitionfunction}, we can choose it so that the partition function is an arbitrary function of the form \eqref{eq:arbitrarypartfn}.

The above analysis leads us to pose the following question, which we will not pursue further in this note.

\begin{question}
    Is there a higher-categorical argument picking out \eqref{eq:partitionfunction} as the preferred $SO(4)$-fixed point data on an object $\mathbf{C}$ of some $4$-category of braided fusion categories?
\end{question}

\subsubsection{Conventions}

All tensor categories are over $\C$.
The symmetric monoidal category of finite-dimensional vector spaces is denoted $\Vect$.

\subsubsection*{Acknowledgements}

 Firstly, I want to thank Pavel Safronov for posing the question answered in this article as well as many interesting conversations about the contents.
 Additionally I thank Arun Debray, Jackson van Dyke, Jin-Cheng Guu, Theo Johnson-Freyd, Andrea Grigoletto , David Jordan, Lukas M\"uller, Noah Snyder and Matthew Yu for useful discussions.
 Comments by Arun Debray, Theo Johnson-Freyd, Cameron Krulewski and Lukas M\"uller on an earlier draft significantly improved the paper.
 I am grateful for the financial support of AARMS and the facilities provided by Dalhousie University to carry out my work.

\section{TQFTs}

\subsection{Locality of TQFTs and the cobordism hypothesis}

Classically, \emph{nonextended TQFTs} are defined as symmetric monoidal functors from the bordism\footnote{Bordism and cobordism are synonyms.} $1$-category $\Bord_{d-1,d}$ to some target symmetric monoidal category $\mathcal{T}$~\cite{atiyah1988topological}.
Here $\Bord_{d-1,d}$ has objects closed oriented\footnote{From now on all manifolds will be assumed to be oriented.} $(d-1)$-dimensional manifolds and morphisms are (diffeomorphism classes relative boundary of) $d$-dimensional bordisms. The monoidal product is given by disjoint union.

To encode full locality, we extend this definition to higher categories.
Recall that there exists an $(\infty,d)$-category $\Bord_d$ with objects $0$-dimensional manifolds, $1$-morphisms $1$-dimensional manifolds with boundary, $2$-morphisms $2$-dimensional manifolds with corners, etc~\cite{calaquescheimbauer}.

\begin{definition}
For $\mathcal{T}$ a symmetric monoidal $d$-category, a \emph{(fully extended) $d$-dimensional TQFT with values in $\mathcal{T}$} is a symmetric monoidal functor
\[
F \colon \Bord_d \to \mathcal{T}
\]
from the (oriented) bordism $d$-category.
Let
$\TQFT_d(\mathcal{T}) =  \Fun_{\mathbb{E}_\infty}(\Bord_d, \mathcal{T})$ denote the space (or $d$-groupoid) of $d$-dimensional TQFTs with values in $\mathcal{T}$.
\end{definition}

\begin{remark}[Geometric structures]
    There are good physical and mathematical reasons to introduce other geometric structures on the bordism category, such as spin structures for fermionic theories or principal $G$-bundles to model background gauge fields.
    However, in this document we focus on oriented TQFTs, just like Atiyah did originally.
\end{remark}

In theory, TQFTs are well-understood in terms of the cobordism hypothesis:

\begin{hypothesis}\cite{baezdolan, luriecobhyp, geometriccobhyp}
    The space $\TQFT(\mathcal{T})$ is homotopy equivalent to $(\mathcal{T}_{fd}^{\cong})^{SO(d)}$.
\end{hypothesis}

We explain the notation in the above.
If $\mathcal{T}$ is a symmetric monoidal $d$-category, $\mathcal{T}_{fd} \subseteq \mathcal{T}$ is the full subcategory on its fully dualizable objects, see \cite[Section 2]{claudiaowen} for an overview of dualizability.
$\mathcal{T}^{\cong}$ denotes the core of $\mathcal{T}$, which is the subcategory obtained by throwing out all non-invertible $k$-morphisms for $k>0$.
The superscript $SO(d)$ denotes the space of homotopy fixed points of an $SO(d)$-action on $\mathcal{T}_{fd}^{\cong}$, whose existence is part of the hypothesis.
The object of $\mathcal{T}_{fd}^{\cong}$ corresponding to a given TQFT is its value on the connected zero-dimensional manifold with its canonical orientation.

For $d<3$, we have a complete rigorous understanding of the $SO(d)$-action and its fixed points~\cite{schommer2009classification, janhesse}.
For $d=3$, there is work in progress by Douglas, Schommer-Pries and Snyder, as sketched in \cite[Section 21]{schommer2014dualizability}, also see \cite{douglas2020dualizable}.
For $d>3$, the $SO(d)$-action is unknown, but we do have various guesses.
Consider for example the $SO(4)$-action on $\BrTens_{fd}^{\cong}$, where $\BrTens$ is the Morita $4$-category of braided tensor categories, also see the next section.
It is known that the induced $SO(2)$-action lifts to the $2$-category of braided tensor categories and its fixed points are balanced braided tensor categories~\cite{salvatore2003framed} (see Section \ref{sec:adjectives} for a review of common definitions for tensor categories, such as balanced and ribbon).
Therefore every balanced braided tensor category in particular defines an $SO(2)$-fixed point in $\BrTens_{fd}^{\cong}$.
Similarly, it is expected that ribbon categories define $SO(3)$-fixed points~\cite[David Jordan's Lecture 5]{simonslectures}.

Note that any extended TQFT induces a nonextended TQFT by forgetting what happens to lower-dimensional manifolds.
Abstractly, the \emph{$(d-1)$-fold loop category} $\Omega^{d-1} \Bord_{d}$, where $\Omega \mathcal{T} = \End_{\mathcal{T}}(1)$, is given by $\Bord_{d-1,d}$.
So if $F\colon\Bord_{d} \to \mathcal{T}$ is a fully extended TQFT, it induces a functor $\Omega^{d-1} F\colon\Bord_{d-1,d} \to \Omega^{d-1} \mathcal{T}$.
Even though it is physically desirable to have a fully extended TQFT, the most immediate physical information---such as partition functions, state spaces and the algebra of local operators---are given in terms of the nonextended TQFT.
Therefore, given a nonextended TQFT $Z$ with values in the symmetric monoidal category $\Vect$ \emph{and} a choice of symmetric monoidal $d$-category $\mathcal{T}$ such that $\Omega^{d-1} \mathcal{T} = \Vect$, it is natural to ask the following question.

\begin{question}
    Does there exist a fully extended TQFT $F$ with values in $\mathcal{T}$ that recovers $Z$?
\end{question}

In that case we say $F$ \emph{extends $Z$ down to points.}
This question---as well as the uniqueness of $F$---is an important guiding principle in the field, see for example \cite{henriques2017chernsimons}.
In practice it is a very difficult and interesting problem to compute the partition function of a TQFT from the object it assigns to a point; we only understand this process fully in dimensions up to two.

\subsection{Target categories}
\label{sec:targets}

Extending down to points strongly depends on the target category; small variations of a target can lead to different dualizable objects, see \cite{carqueville2020extending}.

It is widely accepted that the `correct' target for nonextended TQFTs from a physical perspective is the category $\sVect$ of supervector spaces over $\C$.
The correct target for once extended theories is the Morita $2$-category $\sAlg$ of super algebras,\footnote{Alternatively one can take any variant of $\C$-linear super categories, which all have equivalent subcategory of fully dualizables~\cite[Appendix A]{bartlett2015modular}.} super bimodules and grading-preserving bimodule maps.
Continuing this list (and in particular setting requirements for how to continue the list) is an important guiding problem in the field.
Constructing a target category that is similarly `universal' is work in progress by several authors, see \cite{Teleman_2022} for category level $3$ and \cite{Johnson-Freyd_2023} for the universal $d$-category $\mathcal{U}_d$ of super-duper $d$-vector spaces for $d>3$.

There are however, more well-established target $3$- and $4$-categories.
One such $4$-category is the $4$-category $\BrFus$ of braided fusion categories as defined in \cite[Proposition 3.9]{brochier2021invertible}.
Another construction of higher target categories is an extension of Kapranov and Voevodsky's construction of $2$-vector spaces to arbitrary $d$.
The $d$-category $\Vect_d$ of such $d$-vector spaces is obtained by inductively delooping and adding direct sums.
One issue with defining $\Vect_d$ this way is that it is not idempotent complete in the sense of \cite{gaiotto2019condensations} for $d>2$, but we will not discuss this here.


\begin{remark}[Various notions of tensor category]
Especially for applications to non-semisimple TQFTs, it would be good to generalize from fusion to more general tensor categories.
There is no established terminology for `tensor category' in the literature, as different authors require different finiteness conditions.

One method to construct higher categories of (braided) tensor categories is to take Morita categories of algebras~\cite{claudiathesis} in some sufficiently nice\footnote{The composition of bimodules in $\Alg_{\mathbb{E}_k}(\mathcal{T})$ is a relative tensor product, which can be defined as a coequalizer. This only works if $\mathcal{T}$ has enough colimits, which need to be preserved by the tensor product of $\mathcal{T}$. This is not always the case in relevant examples, e.g.\ a workaround is needed to define $\Alg_{\mathbb{E}_1}(\Rex^{finite})$.} fixed symmetric monoidal $n$-category $\mathcal{T}$.
More specifically, there is a symmetric monoidal $(n+k)$-category $\Alg_{\mathbb{E}_k}(\mathcal{T})$ of which objects are $\mathbb{E}_k$-algebras in $\mathcal{T}$, $1$-morphisms are $\mathbb{E}_{k-1}$-bimodules, and higher morphisms are higher morphisms of $\mathcal{T}$ intertwining the $\mathbb{E}_k$-algebra actions.\footnote{Here we want to define the Morita category using unpointed bimodules, see \cite[Section 3.4.2]{claudiathesis} and \cite[Section 1.7]{claudiaowen} for further discussion.}
For example, $\Alg_{\mathbb{E}_1}(\sVect) = \sAlg$ is the previously mentioned Morita $2$-category of superalgebras.

We define the following versions of the $2$-category of $\C$-linear (i.e.\ enriched and tensored over $\C$), categories:
\begin{itemize}
    \item Let $\Rex$ be the $2$-category of $\C$-linear categories with finite colimits, finite-colimit-completed tensor product and finite colimit preserving linear functors (i.e.\ right exact functors).
\item Let $\Rex^{fin} \subseteq \Rex$ be the full subcategory on the finite abelian categories; those equivalent to $\Mod_A^{fd}$ for some finite-dimensional (not necessarily semisimple) algebra $A$.
\item Let $\Pr$ be the $2$-category of locally presentable $\C$-linear categories and colimit preserving functors.
\end{itemize}
In the literature, authors have defined the $3$-category of tensor categories $\Tens$ as $\Alg_{\mathbb{E}_1}(\Rex^{finite})$, $\Alg_{\mathbb{E}_1}(\Rex)$, or $\Alg_{\mathbb{E}_1}(\Pr)$ in decreasing order of finiteness assumptions.
In order to agree with the terminology of \cite{brochierjordansnyder}, we will define $\Tens = \Alg_{\mathbb{E}_1}(\Pr)$ and $\BrTens = \Alg_{\mathbb{E}_2}(\Pr)$.
In particular, for us tensor categories will in general neither be finite nor rigid, thus diverting from the conventions of the standard reference \cite{egno}.
However, we will mostly work with $\BrFus$, because less is known about invertible objects of $\BrTens$, see \cite[Section 4]{brochier2021invertible}.
\end{remark}

\begin{remark}[Dualizable objects]
Answering the question of what the subcategory of fully dualizable objects are in the above $n$-categories is challenging, see \cite{brochierjordansnyder, douglas2020dualizable} for some results.
In this note however, we are exclusively interested in invertible objects of the target, which are automatically fully dualizable.
\end{remark}

\section{Invertible TQFTs and stable homotopy}
\label{sec:iTQFT}

If $\mathcal{T}$ is a symmetric monoidal $(\infty,d)$-category, let $\Pic \mathcal{T} \subseteq \mathcal{T}$ denote the (non-full) subcategory on objects, $1$-morphisms, \dots, $d$-morphisms which are invertible under compositions and tensor products.
Note that $\Pic \mathcal{T}$ is a full subcategory of the core $\mathcal{T}^{\cong}$, but the inclusion is typically not an equivalence.
A TQFT $\Bord_d \to \mathcal{T}$ is called \emph{invertible} if it factors through $\Pic \mathcal{T}$, i.e.\ if all its values are invertible.
The inclusion of (grouplike $\mathbb{E}_\infty$) $\infty$-groupoids into ($\mathbb{E}_\infty$) $(\infty,d)$-categories admits a left adjoint localization functor that forces all objects and morphisms to become invertible.

Recall that the homotopy hypothesis relates the $(\infty,1)$-category of $\infty$-groupoids to the $(\infty,1)$-category of homotopy types of nice topological spaces.
Similarly, there is a stable homotopy hypothesis which relates $\infty$-groupoids with grouplike $\mathbb{E}_\infty$-structure to a notion from algebraic topology called a connective spectrum.
The stable homotopy hypothesis is closely related to May's recognition principle~\cite{may2006geometry}. One instance of the stable homotopy hypothesis was proven in \cite{moser2022stable}.
We briefly introduce spectra in the next section.

To see how these considerations are useful, first note that the space of functors $\Bord_d \to \Pic \mathcal{T}$ is homotopy equivalent to the space of continuous maps $\| \Bord_d \| \to \Pic \mathcal{T}$.
Here $\|\Bord_d\|$ is the geometric realization of $\Bord_d$, the space corresponding to the $\infty$-groupoid that is the localization of $\Bord_d$ under the homotopy hypothesis.
By a generalization~\cite{schommer2024invertible} of the Galatius-Madsen-Tillmann-Weiss theorem~\cite{gmtw}, the localization of $\Bord_d$ corresponds to the Madsen-Tillmann spectrum $\Sigma^d MTSO(d)$.
\sloppy We are therefore interested in studying the mapping spectrum $\ITQFT(\mathcal{T}) := [\Sigma^d MTSO(d),\Pic \mathcal{T}]$ of $d$-dimensional invertible TQFTs of which the underlying space $\Omega^\infty [\Sigma^d MTSO(d),\Pic \mathcal{T}] \subseteq \TQFT(\mathcal{T})$ is the subspace of invertible TQFTs.\footnote{We refer the reader to \cite[Section 5.2]{freed2021reflection} for a more detailed explanation of this approach.}
To do so, we first collect some results about spectra and maps between them.

\subsection{Spectra}

Spectra are a standard tool in stable homotopy theory that combine homological algebra with topological spaces.
A complete review on spectra is beyond the scope of this note; see \cite[Section 2]{beaudry2018guide} for an introduction.
There is an $(\infty,1)$-category of spectra $\Sp$ with internal hom $[E,F] \in \Sp$ given $E,F \in \Sp$. 
There is a functor from spaces to spectra denoted $\Sigma^\infty_+$ and a right adjoint functor back denoted $\Omega^\infty$.
A spectrum $E$ has homotopy groups $\pi_n(E)$ similar to how spaces do, and they can even be nonzero for negative integers $n$.
A spectrum is called \emph{connective} if $\pi_n(E) = 0$ for $n < 0$.
The functor $\Sigma^\infty_+$ does not preserve homotopy groups, instead giving the \emph{stable} homotopy groups of the space.
The $(\infty,1)$-category of spectra comes with an invertible suspension functor $\Sigma\colon\Sp \to \Sp$ which shifts all homotopy groups by one.
Spectra $E \in \Sp$ give (generalized) cohomology theories in the sense that $h^n(S) := \pi_{-n}[\Sigma^\infty_+ S, E]$ defines a cohomology theory and conversely every cohomology theory is of this form.

Given an abelian group $A$ there is a spectrum $HA$ called the \emph{Eilenberg-Maclane spectrum}, which corresponds to ordinary cohomology with values in $A$.
Just like Eilenberg-Maclane spaces, it is uniquely characterized by the fact that 
\[
\pi_n(HA) =
\begin{cases}
    A & n = 0,
    \\
    0 & n \neq 0.
\end{cases}
\]
However, given another abelian group $A'$, its degree $n$ group cohomology
\[
H^n_{group}(A';A) =  \pi_n [\Sigma^\infty_+ BA', HA] 
\]
is in general different from its \emph{stable cohomology}
\[
H^n_{st}(A';A) =  \pi_n [HA', HA] = \lim_k \pi_n [B^k A', B^k A].
\]
The latter has been computed in low degrees by Eilenberg-Steenrod~\cite{eilenbergSteenrod}.
A useful tool to compute these groups is the iterated bar complex, see \cite[Appendix on the classification of Picard groupoids]{kst}.

\begin{example}
    We have that $H^1_{st}(A;A') \cong \Ext^1(A,A')$.
    The map $H^1_{st}(A;A') \to H^2(BA;A')$ to unstable group cohomology is in general not an isomorphism.
    The former group classifies abelian group extensions, while the latter classifies central extensions.
\end{example}

\subsection{\texorpdfstring{$k$}{k}-invariants and two-term spectra}
\label{sec:2term}

Recall that every space admits a Postnikov tower.
Intuitively, this means it can be constructed iteratively by gluing Eilenberg-Maclane spaces.
The gluing data is given by fibrations, which are a generalization of group extensions to the realm of algebraic topology.
Just like extensions, fibrations with fiber an Eilenberg-Maclane space are classified by a certain cohomology class on the base called a \emph{$k$-invariant}.

This story is entirely similar for spectra.
Going up one level in the Postnikov tower involves a fibration where the fiber is now an Eilenberg-Maclane spectrum, and so now the $k$-invariant will be a stable cohomology class.
We will only need the following variant.
Let $E$ be a spectrum with only $\pi_0$ and $\pi_n$.
Then its connected cover $\pi_{\geq 1} E$ is the Eilenberg Maclane spectrum $\Sigma^n H \pi_n$.
The corresponding fibration
\[
\Sigma^n H \pi_n \to E \to H \pi_0
\]
is classified by a map $H\pi_0 \to \Sigma^{n+1} H \pi_n$.
In other words, such spectra are classified by the triple $(\pi_0, \pi_n, k \in H^{n+1}_{st}(H\pi_0;\pi_n))$.
The following result will be useful to classify invertible TQFTs.
It is strongly inspired by unpublished work of Kreck-Stolz-Teichner~\cite{kst} on the classification of functors between Picard groupoids.

\begin{lemma}
\label{lem:twotermspectra}
    Let $E,F$ be spectra with only $\pi_0$ and $\pi_n$.
    There is a short exact sequence
\[
\adjustbox{scale=.95,center}{%
\begin{tikzcd}[bezier bounding box, column sep = 0] 
        0 \ar[r] & H^n_{st}(\pi_0(E); \pi_n(F)) \ar[r] & {\pi_0[E,F]}  \ar[dll, out=355,in=175, end anchor={[xshift=3ex]}] & \
        \\
    \ &  \{(f_0\colon \pi_0(E) \to \pi_0(F), f_n\colon \pi_n(E) \to \pi_n(F) | f_n \circ k_E = k_F \circ f_0 \in H^{n+1}_{st}(\pi_0(E); \pi_n(F))\} \ar[rr] & \ &  0 
    \end{tikzcd}}
    \]
\end{lemma}
\begin{proof}
    A map $E \to F$ consists of maps $f_0\colon H\pi_0(E) \to H\pi_0(F)$ and $f_n\colon H\pi_n(E) \to H\pi_n(F)$ together with a homotopy filling the square
    \begin{equation}
    \label{eq:homotopy}
    \begin{tikzcd}
        H \pi_0(E) \ar[r,"k_E"] \ar[d,"f_0"] &  \Sigma^{n+1} H \pi_n(E) \ar[d,"f_n"]
        \\
        H\pi_0(F) \ar[ru,Rightarrow,"h"] \ar[r,"k_F"] & \Sigma^{n+1} H \pi_n(F)
    \end{tikzcd}.
    \end{equation}
    The maps $f_0$ and $f_n$ are between Eilenberg-Maclane spectra of the same degree. 
    Since $H^0_{st}(A;B) = \Hom(A,B)$, these are (up to homotopy) equivalent to the data of the homomorphisms $\pi_i(f)\colon\pi_i(E) \to \pi_i(F)$ for $i = 0,n$.
    The homotopy forces the equality $f_n \circ k_E = k_F \circ f_0$ inside 
    \[
    \pi_0 [H \pi_0(E), \Sigma^{n+1} H\pi_n(F)] = H^{n+1}_{st}(\pi_0(E), \pi_n(F)).
    \]
    Given the existence of a homotopy $h$, any other homotopy is given by composing with a self-homotopy of $f_n \circ k_E$.
    By subtracting this map, $h$ is equivalent to a self-homotopy of the zero map, i.e.\ an element of
    \[
    \pi_1 [H\pi_0(E), \Sigma^{n+1} H \pi_n(F)] = \pi_0 [H \pi_0(E), \Sigma^n H \pi_n(F)] = H^n_{st}(\pi_0(E); \pi_n(F)).
    \]
    So homotopy classes of homotopies lying over fixed choices of homotopy classes of maps $[f_0]$ and $[f_n]$ are a torsor over $H^n_{st}(\pi_0(E); \pi_n(F))$.
\end{proof}

\subsection{Picard groupoids of some targets}
\label{sec:Pic(C)}

We discuss Picard groupoids of some of the targets from Section \ref{sec:targets}.

Note that $\Pic \Vect = \Sigma H\C^\times$ since the only tensor invertible object is the one-dimensional vector space which has automorphisms $\C^\times$.
If $\Alg := \Alg_{\mathbb{E}_1}(\Vect)$ is the Morita $2$-category of complex algebras, we also have $\Pic \Alg = \Sigma^2 H\C^\times$.
This follows because $\Omega \Alg = \Vect$ and $\pi_0(\Pic\Alg) = 0$ is the Brauer group of $\C$.\footnote{The Brauer group can be different for other fields. For example, $\Pic \Alg_\R \neq \Sigma^2 H\R^\times$.}

Surprisingly, most targets used for higher-dimensional TQFTs don't have a shift of $H\C^\times$ as their Picard spectrum, the main exception being $\Vect_n$.
It is true that $\Pic \Fus = \Sigma^3 H\C^\times$, but it is an open question whether $\Pic \Tens = \Sigma^3 H\C^\times$, see the discussion in \cite[Section 4]{brochier2021invertible}.
Not just $\pi_1$ but also $\pi_0$ of this spectrum can be nonzero for other fields~\cite{sanford2024invertible}.

In dimension four, the situation is more dire.
It follows by \cite[Theorem 4.2]{brochier2021invertible} that $\pi_0(\BrFus)$ is $\mathcal{W}$, the Witt group of nondegenerate braided fusion categories (see Section \ref{sec:adjectives} for the definitions of nondegenerate and slightly degenerate).
This group is huge and its isomorphism type is a consequence of the results of \cite{davydov2013structure}, as communicated to us by Theo Johnson-Freyd.

\begin{lemma}
There is an isomorphism
    \begin{equation}
    \label{eq:Wittgroup}
    \mathcal{W} \cong \Z/32 \oplus \bigoplus_{\mathbb{N}}( \Z \oplus \Z/2 \oplus \Z/4)
    \end{equation}
\end{lemma}
\begin{proof}
    Consider the Witt group $\sW$ of slightly degenerate braided fusion categories.
    It follows by \cite[Proposition 5.18 and Proposition 5.16]{davydov2013structure} that
    \[
    \sW \cong \bigoplus_{\mathbb{N}} (\Z \oplus \Z/2 \oplus \Z/4).
    \]
    By \cite[Proposition 5.14]{davydov2013structure} the map $\mathcal{C} \mapsto \mathcal{C} \times_{\Vect} \sVect$ defines a homomorphism $\phi:\mathcal{W} \to \sW$ with kernel $\Z/16$.
    It is known that $\mathcal{W}$ has an element of order $32$ 
     and no element of higher order~\cite[Corollary 5.19]{davydov2013structure}.
     It follows by the main result of \cite{minimalnondeg}
 that $\phi$ is surjective, see \cite[Proposition 4.1]{minimalnondeg}.
     By basic extension theory and the fact that there are already countably many copies of $\Z$ we see that the isomorphism type of $\mathcal{W}$ is fixed to be \eqref{eq:Wittgroup}.
\end{proof}

From \cite{brochier2021invertible} we know that $\pi_1(\Pic\BrFus) = \pi_2(\Pic \BrFus) = \pi_3(\Pic \BrFus ) = 0$.
Clearly $\pi_4(\Pic \BrFus) = \C^\times$, so that Lemma \ref{lem:twotermspectra} applies.

Replacing $\BrFus$ with $\BrTens$ makes things much more complicated however.
Not much is known about the map $\pi_0(\Pic \BrFus ) \to \pi_0(\Pic \BrTens)$. 
It is easy to construct Morita invertible braided tensor categories which are Morita equivalent but not isomorphic to a braided fusion category.
Clearly $\pi_3(\Pic \BrTens)$ is zero and $\pi_4(\Pic \BrTens) = \C^\times$.
Its $\pi_2$ consists of equivalence classes of invertible objects of $\Pr$, which has recently been shown to be zero as well~\cite{stefanich2023classification}.
$\pi_1(\Pic \BrTens )$ is conjectured to be zero. 

The universal target $\mathcal{U}_d$ of super-duper vector spaces in dimension $d$ has $\Pic \mathcal{U}_d  = \pi_{\geq 0} \Sigma^d I\C^\times$.
Here $I\C^\times$ denotes the Brown-Comenetz dual of the sphere spectrum, characterized by the universal property that
\[
\pi_0[E, I\C^\times] = \Hom(\pi_0(E), \C^\times)
\]
for all spectra $E$.
In fact, this requirement was one original desideratum for this $d$-category.
In particular, $\Pic \sVect = \pi_{\geq 0} \Sigma I\C^\times$ and $\Pic \sAlg = \pi_{\geq 0} \Sigma^2 I\C^\times$, see \cite{freed2012lectures, freed2014anomalies}.

\subsection{The classification of invertible TQFTs}

In this section, we will apply the results of Section \ref{sec:2term} to study invertible TQFTs.

    Recall that in the fully extended setting, we can replace $\Bord_d$ with $\Sigma^d MTSO(d)$. 
    Its homotopy groups are known to be the (oriented) bordism groups $\pi_j(\Sigma^d MTSO(d)) = \Omega_j$ for $j<d$, which are completely understood. 
    
    However, $\pi_d (\Sigma^d MTSO(d))$ is not quite isomorphic to $\Omega_d$.
    Instead, it is a Reinhart bordism or SKK group $SKK_d$~\cite{reinhart1963cobordism, skbook, MR3212580}.
        It is still true that $SKK_d$ is a quotient of the Grothendieck group of the monoid of $d$-dimensional closed manifolds, but the relation is weaker than bordism.
    It is known that~\cite{skbook}
    \[
    SKK_d \cong
    \begin{cases}
    \Omega_d \times \Z & d \equiv 0 \pmod 2
    \\
        \Omega_d \times \Z/2 & d \equiv 1 \pmod 4
        \\
        \Omega_d & d \equiv 3 \pmod 4
    \end{cases}.
    \]
    Moreover, if $d \equiv 0 \pmod 4$, the projection $SKK_d \to \Z$ on the second factor is given by 
    \[
    M \mapsto \frac{\sigma(M) - \chi(M)}{2}.
    \]
We thus know the homotopy groups of $\Sigma^d MTSO(d)$ up to degree $d$, also see \cite[Section 7.2]{schommer2024invertible}.

    Even though $\Sigma^d MTSO(d)$ potentially has many more nontrivial homotopy groups above degree $d$, they will not matter in this discussion as long as our target $\mathcal{T}$ is a $d$-category, and not an $(\infty,d)$-category; the truncation functor $\pi_{\leq d}$ sits in an adjunction which assures that 
    \[
    [\Sigma^d MTSO(d),\Pic \mathcal{T}] = [\pi_{\leq d} \Sigma^d MTSO(d), \Pic \mathcal{T}]
    \]
    if $\Pic \mathcal{T}$ has trivial homotopy groups above degree $d$.
    In other words: the TQFT factors through the homotopy $d$-category of the bordism $(\infty,d)$-category $\Bord_d$, and so we can replace $\Sigma^d MTSO(d)$ by $E_d := \pi_{\leq d} \Sigma^d MTSO(d)$.

The approach using spectra also applies to nonextended invertible TQFTs because $\|\Bord_{d-1,d}\| = \pi_{\geq 0} \Sigma^1 MTSO(d)$~\cite{gmtw}.
Let $\ITQFT_{d-1,d}(\mathcal{T}) = [\pi_{\geq 0} \Sigma^1 MTSO(d), \Pic \mathcal{T}]$ be the spectrum of nonextended invertible TQFTs.
These are maps of two-term spectra, which are completely understood using Picard groupoids.
It is therefore not difficult to do computations with nonextended invertible TQFTs, independently of the target category. 
For example:

\begin{proposition}
    We have 
    \[
    \pi_0 \ITQFT_{d-1,d}(\Vect) \cong \{Z\colon SKK_d \to \C^\times : Z(Y \times S^1) = 1 \},
    \]
    where $Z$ is a homomorphism and $\pi_0 \ITQFT_{d-1,d}(\Vect)$ is the group of equivalence classes of nonextended invertible TQFTs.
\end{proposition}
\begin{proof}
We can replace $E_d = \pi_{\leq d} \Sigma^d MTSO(d)$ by a further connected cover to obtain a Picard groupoid with just two homotopy groups $\Omega_{d-1}$ and $SKK_d$.
This two-term spectrum has a $k$-invariant in
\[
H^2_{st}(SKK_d; \Omega_{d-1}) \cong \Hom\left(\frac{SKK_d}{2SKK_d}, \Omega_{d-1}\right).
\]
By the correspondence between Picard groupoids and stable $1$-types~\cite{sinhthesis, johnson2012modeling}, it is given by $Y \mapsto Y \times S^1$.
We now apply Lemma \ref{lem:twotermspectra} for $n=1$.
    We have that 
    \[
    H^1_{st}(\Omega_{d-1},\C^\times) = \Ext^1(\Omega_{d-1}, \C^\times) =0
    \]since $\C^\times$ is injective, so the $k$-invariant obstruction always vanishes.
    Since $\pi_0 \Pic \Vect = 0$, we obtain the result.
\end{proof}

\begin{example}
    Since $SKK_1 \cong \Z/2$ is generated by $S^1$, we have that $\ITQFT_{0,1}(\Vect) = 0$.
    On the other hand, because $\mathcal{U}_1 = \sVect$ satisfies $\Pic \sVect  = \pi_{\geq 0} \Sigma I\C^\times$, we have $\ITQFT_{0,1}(\sVect) = \Hom(SKK_1,\C^\times) \cong \Z/2$.
    This is an example of a non-unitary Kervaire TQFT, see \cite[Example 6.6]{simonareneeskk}.
\end{example}

\subsection{\texorpdfstring{$4d$}{4d} invertible TQFTs}
\label{sec:4dTQFTs}

We now restrict to dimension $4$.
    We have that $\Omega_0 \cong \Z$ is given by sending the positively oriented point to $1$ and the negatively oriented point to $-1$, while $\Omega_1 = \Omega_2 = \Omega_3 = 0$.
    The signature defines an isomorphism $\Omega_4 \cong \Z$, so that in particular $SKK_4 \cong \Z \times \Z$ by the Euler characteristic and the signature, which satisfy the single relation that they are equal modulo two.
In particular, we obtain
\[
\pi_0 \ITQFT_{3,4}(\Vect) \cong \pi_0 \ITQFT_{3,4}(\sVect) \cong \C^\times \times \C^\times.
\]
The TQFT corresponding to $\lambda_1, \lambda_2 \in \C^\times$ has partition function 
\[
Z_{(\lambda_1, \lambda_2)}(X^4) = \lambda_1^{\sigma(X)} \lambda_2^{\frac{\chi(X)-\sigma(X)}{2}}.
\]
We would like to understand for several targets whether such theories extend to a fully extended invertible TQFT.
To make sure we can talk about recovering $Z_{(\lambda_1, \lambda_2)}$ on top-dimensional manifolds, we make the following definition.

\begin{definition}
    Let $\mathcal{T}$ be symmetric monoidal $4$-category.
    We call $\mathcal{T}$ \emph{top-complex} if $\Aut \id_{\id_{\id_1}} = \C^\times$.
\end{definition}

This is always the case for target categories for which partition functions take values in the complex numbers.
In particular, this holds for all targets $\mathcal{T}$ for which nonextended theories are the same as TQFTs with target $\Vect$ or $\sVect$.

\begin{theorem}
\label{th:TQFTcomputation}
    Let $\mathcal{T}$ be a top-complex symmetric monoidal $4$-category such that $\Pic \mathcal{T} $ has no $\pi_1, \pi_2$ and $\pi_3$.
    Then there is an exact sequence
    \[
    0 \to \Z/6 \to \pi_0  \ITQFT_4(\mathcal{T}) \to \C^\times \times \C^\times \times \pi_0 (\Pic \mathcal{T}) \to 0.
    \]
\end{theorem}
\begin{proof}
    We see that Lemma \ref{lem:twotermspectra} applies to $E := E_4 = \pi_{\leq 4} \Sigma^4 MTSO(4)$ and $F = \Pic \mathcal{T} $ for the case $n = 4$.
    It is known that $H^5_{st}(A; \C^\times) = \Hom(\tors_2(A) \oplus \tors_3(A), \C^\times)$, where $\tors_p(A)$ is the $p$-torsion subgroup~\cite{eilenbergSteenrod}.
    It follows that the obstruction group $H^5_{st}(\Z; \C^\times)$ in which $f_n \circ k_E$ and $k_F \circ f_0$ live is zero.
    By the universal coefficient theorem,
    \[
    H^4_{st}(\pi_0(E); \C^\times) = \Hom(H_4^{st}(\Z; \Z), \C^\times) \cong \Z/6
    \]
    using that $H_4^{st}(\Z; \Z) \cong \Z/6$.
    Finally we use that $\pi_4 E = \Z \times \Z$ as in the nonextended case to obtain the two $\C^\times$ terms.
\end{proof}

\begin{corollary}
\label{cor:BrfusTQFTclassification}
    There is an exact sequence
    \[
    0 \to \Z/6 \to \pi_0  \ITQFT_4(\BrFus) \to \C^\times \times \C^\times \times \mathcal{W} \to 0.
    \]
\end{corollary}

We have thus proven Theorem \ref{mainth}.

\begin{remark}
    Even though the relevant obstruction group $H^5_{st}(\Z; \C^\times)$ in Theorem \ref{th:TQFTcomputation} is always zero, the groups 
    \[
    H^5_{st}(\Z; \Z \times \Z) \cong \Z/6 \times \Z/6 \quad \text{ and } \quad H^5_{st}(\mathcal{W}; \C^\times) \cong \bigoplus_{\mathbb{N}} \Z/2
    \]
    in which the $k$-invariants of the spectra $E_4$ and $\Pic \BrFus$ live are nonzero.
    We expect the computation of these $k$-invariants to be relevant for deriving the isomorphism class of the extension in Corollary \ref{cor:BrfusTQFTclassification}.
\end{remark}

\begin{example}
Suppose $\mathcal{T} = \Vect_4$ so that $\Pic \mathcal{T} = \Sigma^4 H\C^\times$.
Then the above theorem recovers the first part of \cite[Theorem 7.6.3]{schommer2024invertible}. 
This target has also been used to discuss invertible Crane--Yetter theories in \cite[Section 1.3]{schommer2018tori}.
\end{example}

\begin{remark}[Obstruction theory for spaces vs spectra]
    Let $F$ be a spectrum with only $\pi_0$ and $\pi_4$. 
    It is tempting to attempt to build a map $MTSO(4) \to F$ by picking a point $y\in F$ and extending $\Z = \pi_0(\Sigma^4 MTSO(4)) \to \pi_0 (F) \ni y$ to a map into $F$ by obstruction theory.
    In other words, we just restrict to the component of $\Omega^\infty F$ corresponding to $y$ and use that this is a $B^4 \C^\times$.
    For example if $F$ is top-complex, we know that $H^4(\Omega^\infty \Sigma^4 MTSO(4); \pi_4(F)) = \C^\times \times \C^\times$, so naively there seems to be exactly one TQFT for every element of $\pi_0(F)$ and every element of $\C^\times \times \C^\times$.
    The problem with this reasoning in general is that this only gives a map of spaces, not of spectra.
    Therefore it does construct a functor from the bordism category to the target, but it is not symmetric monoidal.
    The symmetric monoidality of a functor is data, and Theorem \ref{th:TQFTcomputation} shows this data always exists and is classified up to symmetric monoidal natural isomorphism by the group $\Z/6$.
\end{remark}

\subsection{Analogy with 2d}

There is an analogue of Theorem \ref{th:TQFTcomputation} for two-dimensional invertible TQFTs.
Indeed, $\pi_{\leq 2} \Sigma^2 MTSO(2)$ has only $\pi_0 = \Z$ and $\pi_2 = \Z$ so that the proof applies with two key differences:
    \begin{itemize}
        \item Because $H^2_{st}(\Z;\C^\times) \cong \Z/2$, the first factor in the short exact sequence is a $\Z/2$ instead of $\Z/6$. 
        \item Because $\pi_2(\Sigma^2 MTSO(2)) = \Z$, there is only one factor of $\C^\times$ instead of two.
    \end{itemize}
        For example, suppose the target is the $2$-category $\Alg$ of complex algebras. 
    Since the Picard is $\Sigma^2 H\C^\times$, we get a short exact sequence
    \[
    0 \to \Z/2 \to \pi_0  \ITQFT_2(\Alg) \to \C^\times \to 0.
    \]
    It turns out that this exact sequence does not split.
    This can be seen explicitly by recalling that an extended 2d TQFT is given by a semi-simple finite-dimensional symmetric Frobenius algebra.
    A symmetric Frobenius structure on the only Morita-invertible algebra (up to Morita equivalence) $\C$ is classified by a single nonzero complex number $\lambda \in \C^\times$, the value of the trace at $1$.
    The partition function corresponding to $\lambda$ on a closed surface $\Sigma$ is given by $\lambda^{\chi(\Sigma)}$, see \cite[Example 1.2]{schommer2024invertible}.
    This number only depends on $\lambda^2$, since the Euler characteristic is even for closed oriented surfaces.
    In particular, there are two isomorphism classes of TQFTs that extend a given two-dimensional nonextended invertible TQFT.


    Note that confusingly for the universal target $\mathcal{U}_2 = \sAlg_\C$ of superalgebras, the fully extended TQFT should instead be determined by the partition function.
    This follows by the universal property of the Brown-Comenetz dual since $\Pic \sAlg = \pi_{\geq 0} \Sigma^2 I\C^\times$.
    The solution to this paradox is that one can show that the symmetric Frobenius algebra $(\C,\lambda)$ is Morita equivalent to $(\C,- \lambda)$ inside $\sAlg$ via the invertible $(\C,\C)$-bimodule given by the odd line.

Separately, note that for unoriented invertible TQFTs, the partition function $\lambda^{\chi(\Sigma)}$ depends on more than just $\lambda^2$.
This corresponds to the fact that the map 
\[
\Z \cong SKK_2 \cong \pi_2 \Sigma^2 MTSO(2) \to \pi_2 \Sigma^2 MTO(2) \cong SKK_2^O \cong \Z
\]
given by comparing unoriented and oriented bordism is given by multiplication by two.
We now briefly dwell on the question of how this square and the previously discussed square combine.

Since the unoriented bordism group $\Omega^O_1$ vanishes, Lemma \ref{lem:twotermspectra} applies to maps from $\Sigma^2MTO(2)$ to $\Pic \Alg$.
Because $\Omega_0^O = \Z/2$ and pulling back along the mod two map is an isomorphism 
\[
\Z/2 \cong H^3_{st}(\Z/2;\C^\times)  \to H^3_{st}(\Z;\C^\times) \cong \Z/2,
\]
we see that extending downwards requires yet another $\Z/2$ in the unoriented case.

We now show the somewhat surprising fact that this further extension is split.

\begin{lemma}
    The $k$-invariant of $\pi_{\leq 2} \Sigma^2 MTO(2)$ is trivial.
\end{lemma}
\begin{proof}
Note that the $k$-invariant lives in $H^3_{st}(\Z/2; \Z) \cong \Z/2$ and so it is not immediate that it vanishes.
We apply Lemma \ref{lem:twotermspectra} to $E = \pi_{\leq 2} \Sigma^2 MTSO(2)$ and $F = \pi_{\leq 2} \Sigma^2 MTO(2)$ and use the fact that we know a map $\Sigma^2 MTSO(2) \to \Sigma^2 MTO(2)$.
Note that
\[
\Z \cong \Omega^{SO}_0 \cong \pi_0 \Sigma^2 MTSO(2) \to \pi_0 \Sigma^2 MTO(2) \cong \Z/2
\]
is surjective and so induces an isomorphism 
\[
\Z/2 \cong H^3_{st}(\Z/2;\Z) \xrightarrow{\pmod 2 ^*} H^3_{st}(\Z;\Z) \cong \Z/2.
\]
Separately, multiplication by two induces the zero map on $H^3_{st}(\Z;\Z) \cong \Z/2$.
We see that the $k$-invariant obstruction vanishes if and only if the $k$-invariant of $\pi_{\leq 2} \Sigma^2 MTO(2)$ itself vanishes in $H^3_{st}(\Z/2; \Z)$.
\end{proof}

The situation can therefore be summarized in the diagram
\[
\begin{tikzcd}[column sep = 50]
    \C^\times \ar[r, "(.)^2"] \ar[d, equals] & \C^\times \ar[d, equals]
    \\
    \pi_0 \ITQFT(\Alg) \ar[r] & \pi_0 \ITQFT_{1,2}(\Alg)
    \\
    \pi_0 \ITQFT^O(\Alg) \ar[r] \ar[u] \ar[d, equals] & \pi_0 \ITQFT_{1,2}^O(\Alg) \ar[u] \ar[d, equals]
    \\
    \C^\times \times \{\pm \} \ar[r,"{(\lambda,\sigma) \mapsto \lambda \sigma}"] \ar[uuu, bend left=70,"{(\lambda,\sigma) \mapsto \lambda}"] &  \C^\times \ar[uuu,bend right=100,"(.)^2",swap]
\end{tikzcd} \quad ,
\]
where $\ITQFT^O$ denotes the spectrum of unoriented extended two-dimensional invertible TQFTs and $\{\pm\}$ is the group $\Z/2$ written multiplicatively.
Note that this is a pullback diagram in groups with both kernels being $\Z/2$; the pullback of the nontrivial group extension of $\C^\times$ by $\Z/2$ under the squaring map on $\C^\times$ is the trivial extension.

The number $\sigma \in \{\pm\}$ has an algebraic interpretation as classifying the structure of a stellar algebra~\cite[Section 3.8.6]{schommer2009classification} on the algebra $\C$ assigned to the point.
This stellar structure does not come from a $*$-algebra structure on $\C$.
However, it can be realized as the $*$-structure on the Morita-equivalent algebra $M_2(\C)$ corresponding to the real structure retrieving the quaternions.

To finish the discussion, we also briefly mention the universal target of superalgebras in the unoriented case.
First note that taking the odd line as a stellar structure is not allowed because it is not compatible with the Frobenius structure in the sense of \cite[Definition 5.22]{Luukas}, so in $\sAlg$ we still only have the stellar structures parametrized by $(\lambda, \sigma)$.
Now the same trick using the odd line as in the oriented case now shows that $(\lambda, \sigma)$ and $(- \lambda, -\sigma)$ are equivalent.
Note how we could have arrived at the same conclusion by applying the universal property of $I\C^\times$ to the above diagram.

\section{Crane--Yetter}

\subsection{Some adjectives for tensor categories}
\label{sec:adjectives}

A balanced tensor category is a braided tensor category $\mathcal{C}$ with a balancing.
A balancing is defined as a monoidal trivialization of the monoidal functor given by the identity with monoidal data $\beta_{y,x} \beta_{x,y}$, i.e.\ a natural automorphism $\theta$ of the identity functor satisfying
\[
\theta_{x \otimes y} = \beta_{y,x} \beta_{x,y} (\theta_x \otimes \theta_y)
\]
for all $x,y \in \mathcal{C}$.
Ribbon categories (also called tortile categories) are rigid balanced tensor categories in which the balancing (also called the twist) satisfies $\theta_{x^*} = \theta_x^*$. Ribbon fusion categories are also called premodular categories. After a choice of left versus right twisting, a balancing on a rigid braided tensor category is equivalent to a pivotal structure. For rigid braided fusion categories, this pivotal structure is spherical if and only if $\theta_{x^*} = \theta_x^*$. In particular, a premodular category is equivalent to a spherical braided fusion category. See \cite[Appendix A.2]{henriques2016categorified} for a survey. 

A spherical braided fusion category $\mathcal{C}$ is called \emph{nondegenerate} if its S-matrix with entries $\Tr (\beta_{x,y} \beta_{y,x}) \in \C$ for $x,y \in \mathcal{C}$ simple is nondegenerate~\cite[Definition 2.27]{drinfeld2010braided}.
This happens if and only if $\mathcal{C}$ is invertible in $\BrTens$~\cite[Theorem 1.1]{brochier2021invertible} and if and only if it is invertible in $\BrFus$~\cite[Theorem 3.20]{brochier2021invertible}, so that being nondegenerate is independent of the balancing.
Being nondegenerate is also equivalent to $Z_{\mathbb{E}_3}(\mathcal{C}) \cong \Vect$ where $Z_{\mathbb{E}_3}$ denotes the M\"uger center~\cite[Proposition 3.7]{drinfeld2010braided}.
We say $\mathcal{C}$ is \emph{slightly degenerate} if $Z_{\mathbb{E}_3}(\mathcal{C}) \cong \sVect$.
If $\mathcal{C}$ is a fusion category, then $Z_{\mathbb{E}_2}(\mathcal{C})$ is nondegenerate~\cite[Corollary 3.9]{drinfeld2010braided}, where $Z_{\mathbb{E}_2}$ is the Drinfeld center.
A premodular category is called modular if it is nondegenerate.
If $\mathcal{C}$ is a spherical fusion category, then $Z_{\mathbb{E}_2}(\mathcal{C})$ is modular.

\subsection{Crane--Yetter as a nonextended TQFT}

Crane--Yetter is a four-dimensional TQFT originally defined in \cite{craneyetter} as a state-sum model associated to a given modular tensor category $\mathcal{C}$, see \cite{craneyetterkauffman} for details and the generalization to ribbon fusion categories.
It is well-known that Crane--Yetter is an invertible TQFT if and only if the ribbon fusion category is modular, so we will restrict to that case to make sure the arguments of the previous sections apply.
This result is surprisingly independent of the target category and of its extension beyond a once extended TQFT~\cite{schommer2018tori}. 
Separately, it is a fact that finite ribbon categories $\mathcal{C}$ which are not nondegenerate are not invertible in $\BrTens$, so fully extended TQFTs that assign $\mathcal{C}$ to a point can't be invertible unless $\mathcal{C}$ is nondegenerate~\cite{brochier2021invertible}.

As an invertible TQFT, Crane--Yetter has the potential to be an anomaly of a three-dimensional TQFT \cite{freed2014anomalies, freed2014relative}.
Indeed, Reshitikhin-Turaev~\cite{reshetikhinturaev} for $\mathcal{C}$ is a boundary theory for Crane--Yetter~\cite{barrett2007observables}.
This therefore connects it also to Chern-Simons theory, where $\mathcal{C}$ is related to the representation category of the quantum group.
Crane--Yetter is also related to BF-theory~\cite{baezBF}.

The partition function for a given modular fusion category $\mathcal{C}$ on a four-dimensional closed manifold $X$ is given by~\cite{crane1993evaluating}
\begin{equation}
    \label{eq:CYpartition}
Z(X) =  (\dim \mathcal{C})^{\chi(X)/2} e^{2\pi i c\sigma(X)/8}.
\end{equation}

We will now briefly get into the question: for which invertible nonextended TQFTs corresponding to elements of $\C^\times \times \C^\times$ does there exist a modular tensor category $\mathcal{C}$ realizing it?
First recall that the global dimension of $\mathcal{C}$ is a positive real number~\cite[Theorem 2.3]{etingof2005fusioncats}, so we have a preferred square root in the above equation.
Not everything is known about what numbers can be the global dimension of a fusion category.
However, it is known that 
\begin{enumerate}
    \item $\dim \mathcal{C}$ is an algebraic number,
    \item $\dim \mathcal{C} \geq 1$ and equal to $1$ iff $\mathcal{C} = \Vect$,
    \item if $\lambda$ is in the image of $\dim$, then so is $f(\lambda)$ for $f$ any automorphism of the field $\C$,
    \item $\dim \mathcal{C}$ is totally positive, which means that $f(\dim \mathcal{C}) > 0$ for every automorphism of the field $\C$,
\end{enumerate}
see \cite{ostrik2018globaldim}.
Modular fusion categories are in particular spherical, for which there are further restrictions as explained in loc.\ cit.
For example, in that case the first allowed global dimension larger than $1$ is $\frac{5-\sqrt{5}}{2}$.

The central charge is an element of $U(1)$ by definition, and it turns out to be a root of unity. 
It depends strongly on the choice of modular structure, to the point where it is known that any root of unity can be realized.

There is moreover a subtle interplay between the two $\C^\times$ factors. 
For example, \cite[Proposition 3.4]{etingof2005fusioncats}
gives a lower bound for $\dim \mathcal{C}$ in terms of the central charge.

The global dimension is also not fixed in a Witt class, so that it does not define a homomorphism $\mathcal{W} \to \C^\times$.
For example, the global dimension of $Z_{\mathbb{E}_2}(\mathcal{C})$ is $(\dim \mathcal{C})^2$ \cite[Theorem 2.15]{etingof2005fusioncats}.
Pseudounitary nondegenerate braided fusion categories on the other hand have a canonical modular structure and therefore a central charge.
They form a subgroup $\mathcal{W}_{un} \subseteq \mathcal{W}$.
This gives a homomorphism $\mathcal{W}_{un} \to \Q/8\Z$~\cite[Corollary 5.28]{davydov2013witt}.

\subsection{Extending Crane--Yetter to a point}

Crane--Yetter is often described as a fully extended field theory, even though a complete published treatment has not appeared to our knowledge. 
The result seems to be claimed first by Dan Freed in joint work with Constantin Teleman as he shared in several talks \cite{432876, danlens}. 
He also mentions the relationship between taking $SO(4)$-fixed points and fixing the partition function of Crane--Yetter we have proven here.
Kevin Walker also has unpublished work, some of which can be found online, such as the slides \cite{walkerpremodular} and the notes \cite{walkertqft}, which have been applied to topological insulators in \cite{walkerwang}.

There are also several more detailed, explicit and published works in this direction and the main tools are usually skein theory~\cite{roberts1995skein} and factorization homology~\cite{kirillov2022factorization}. 
The references \cite{tham2021category, jinthesis, cookethesis} have partial results on extending Crane--Yetter to a $4,3,2$-theory with values in a certain $2$-category of $\C$-linear categories, and a full proof was recently given in \cite[Theorem 4.2]{haioun2025non}. 
We have also been made aware of work on progress by Jin-Cheng Guu to extend it to a $4,3,2,1$-theory.
Unfortunately, none of these references prove the full extendedness claimed by Freed-Teleman and Walker.

The following is an immediate implication of Corollary \ref{cor:BrfusTQFTclassification} for Crane--Yetter.

\begin{theorem}
    Let $\mathcal{C}$ be a modular fusion category. There are six equivalence classes of TQFTs with target $\BrFus$ which both assign $\mathcal{C}$ to the positively oriented point and have partition function \eqref{eq:CYpartition}.
\end{theorem}

\begin{remark}[Crane--Yetter is reflection-positive]
Atiyah defined reflection-positive (or unitary) TQFTs for nonextended non-invertible TQFTs with target $\Vect$. 
Freed-Hopkins~\cite{freed2021reflection} have defined reflection-positivity for extended invertible TQFTs with target $I\C^\times$ in a way that agrees for nonextended invertible TQFTs with target super Hilbert spaces.
They also proved that a given partition function $Z\colon\pi_n \Sigma^n MTSO(n) \to \C^\times$ extends (uniquely) to a nonextended reflection-positive ITQFT if and only if $Z(S^n)$ is a positive real number and $Z/Z(S^n)$ is $U(1)$-valued.
We see that Crane--Yetter is a reflection-positive TQFT in the nonextended sense.
By \cite[Theorem 8.29]{freed2021reflection}, this is true as well for reflection-positivity in the invertible extended sense with target $I\C^\times$.
We expect that there is a dagger $4$-category~\cite{higherdagger} of unitary braided fusion categories so that reflection positive theories (i.e.\ higher dagger functors) with this target are given by unitary modular fusion categories.
\end{remark}

\section{Extra topics}

\subsection{Partial \texorpdfstring{$SO$}{SO}-fixed point data}

In this section we consider the mapping spectrum 
\[
\ITQFT^{SO(k)}(\mathcal{T}) = [\pi_{\geq 0} \Sigma^4 MTSO(k), \Pic \mathcal{T}] 
\]
for $k<4$.
In terms of invertible TQFTs, such maps can be understood as categorified TQFTs.
If $k = 3$ for example, a map 
\[
\pi_{\geq 0} \Sigma^4 MTSO(3) \to \Pic \mathcal{T}
\]
for a symmetric monoidal $4$-category $\mathcal{T}$ corresponds to an invertible symmetric monoidal functor $\Bord_3 \to \mathcal{T}$.
Here we view $\Bord_3$ as a $(4,3)$-category, where the $4$-morphisms are diffeomorphisms.
Alternatively, such a map corresponds to a four-dimensional fully extended invertible TQFT with $SO(3)$-structure, i.e.\ where the four-dimensional bordisms come equipped with a combing in the time direction.

Starting with $k=1$ corresponding to framed theories, we have that $\Sigma MTSO(1) = \mathbb{S}$ is the sphere spectrum $\mathbb{S} := \Sigma^\infty_+(\pt)$.
The framed cobordism hypothesis reduces to the fact $[\mathbb{S}, X] \cong X$. 
In particular, for every non-degenerate braided tensor category there exists a non-trivial fully extended invertible 4d framed
TQFT, which is sometimes regarded as `framed Crane--Yetter' \cite{jackson}\footnote{More generally, any braided fusion category gives a four-dimensional framed TQFT by the cobordism hypothesis, since \cite{brochierjordansnyder} prove that all of $\BrFus$ is fully dualizable.}.
Confusingly however, the partition function of a framed four-dimensional invertible TQFT is necessarily trivial because the fourth stable stem $\pi_4(\mathbb{S})$ vanishes. 
This is not surprising since every framed manifold has trivial signature and Euler characteristic.
This confusion is called the \emph{Crane--Yetter paradox}~\cite{simonslectures}.

We have already seen how to compute $\pi_{i} (\Sigma^4 MTSO(k))$ for $i \leq k$ using bordism and SKK groups.
It is also possible to compute the remaining homotopy groups $\pi_4(\Sigma^3 MTSO(3))$, $\pi_3(\Sigma^2 MTSO(2))$ and $\pi_4(\Sigma^2 MTSO(2))$ using spectral sequence methods, see Table \ref{tab:partial_fixpts} for the combined results.

\begin{table}[h]
\centering
\begin{tabular}{l|lllll}
 & 0 & 1 & 2 & 3 & 4 \\
 \hline
1 & $\Z$  & $\Z/2\Z$ & $\Z/2\Z$ & $\Z/24\Z$ & $0$ \\
2 & $\Z$ & $0$ & $\Z$ & $0$ & $\Z$ \\
3 & $\Z$ & $0$ & $0$ & $0$ & $\Z$ \\
4 & $\Z$ & $0$ & $0$ & $0$ & $\Z \oplus \Z$
\end{tabular}
\caption{Homotopy groups of $\Sigma^k MTSO(k)$}
\label{tab:partial_fixpts}
\end{table}

Let $\mathcal{T}$ be a target $4$-category satisfying the assumptions of Theorem \ref{th:TQFTcomputation}.
For $k = 3$, it follows that the group of fully extended four-dimensional invertible TQFTs with $SO(3)$-structure fits into a short exact sequence
\[
0 \to \Z/6 \to \pi_0 \ITQFT^{SO(3)}(\mathcal{T}) \to \pi_0 (\Pic \mathcal{T}) \times \C^\times \to 0,
\]
since $\pi_4 (\Sigma^4 MTSO(3)) = \Z$. 
Consider the Genauer sequence~\cite{genauer2012cobordism}
\begin{align*}
\dots&\to \pi_1 ( \Sigma^\infty_+ BSO(4)) \to \pi_1(MTSO(4)) \to \pi_0 (MTSO(3)) \xrightarrow{\chi} \Z \cong \pi_0 (\Sigma^\infty_+ BSO(4)) \to 
\\
& \to \pi_0 (MTSO(3)) = 0 \to \dots
\end{align*}
Note that $\pi_1 (\Sigma^\infty_+ BSO(4)) = 0$ by Hurewicz theorem since $H_1(BSO(4)) =  0$, so it follows that the map 
\[
\Z \cong \pi_4(\Sigma^3 MTSO(3)) \to \pi_4(\Sigma^4 MTSO(4)) \cong \Z \times \Z
\]
is given by the inclusion of the kernel of the Euler characteristic map $\pi_4(\Sigma^4 MTSO(4)) \to \Z$.
We conclude that the map $\pi_0 \ITQFT^{SO(4)}(\mathcal{T}) \to \pi_0 \ITQFT^{SO(3)}(\mathcal{T}) \cong \C^\times$ is given by evaluation on a class with signature $2$ and Euler characteristic $0$.
We have thus shown

\begin{theorem}
    The group of once categorified fully extended invertible TQFTs $\Bord_3 \to \BrFus$ is a $\Z/6$-extension of $\C^\times \times \mathcal{W}$.
    Extensions upwards to a functor $\Bord_4 \to \BrFus$ are classified by $\C^\times$.
\end{theorem}

It is expected that algebraically this extra $\C^\times$ factor is described by a modified trace on $\mathcal{C}$, see \cite{costantino2023skein} for a result of this type in the non-semisimple setting.

\begin{remark}[$2$-dimensional twice categorified TQFTs]
Theorem \ref{th:TQFTcomputation} does not apply to the case $k=2$ because $\pi_2(\Sigma^2 MTSO(2)) \neq 0$.
It would be interesting to find another route to compute $\pi_0 \ITQFT^{SO(2)}(\mathcal{T})$. 
 Note that the theorem also does not apply to $k = 1$, but we have $\pi_0 \ITQFT^{SO(1)}(\mathcal{T}) = \pi_0(
\Pic \mathcal{T})$ as mentioned before.
\end{remark}

It would be interesting to understand more explicitly how the $SO(4)$-fixed point data on a choice of $[\mathcal{C}] \in \mathcal{W}$ is related to a choice of ribbon structure on $\mathcal{C}$.
For example, note that the global dimension of $\mathcal{C}$ does not depend on the ribbon structure and so it is not clear how hypothetical $SO(4)$-fixed point data induced by the categorical structure on $\mathcal{C}$ gives the correct element of $\C^\times \times \C^\times$.

\subsection{Partially extended theories}


Partially extended invertible TQFTs are classified by maps out of $\pi_{\geq k} \Sigma^4 MTSO(4)$.
Note that 
\[
\pi_{\geq 1} \Sigma^4 MTSO(4) \cong \pi_{\geq 2} \Sigma^4 MTSO(4) \cong \pi_{\geq 3} \Sigma^4 MTSO(4) \cong \pi_{\geq 4} \Sigma^4 MTSO(4).
\]
We thus obtain that the group of partially extended invertible TQFTs with target $\mathcal{T}$ satisfying the assumptions of Theorem \ref{th:TQFTcomputation} is isomorphic to the group of nonextended invertible TQFTs.

We also see that the map 
\[
\pi_0 \ITQFT(\mathcal{T}) \to \pi_0 \ITQFT_{3,4}(\mathcal{T}) \cong \C^\times \times \C^\times
\]
is given by the obvious projection in the short exact sequence of Theorem \ref{th:TQFTcomputation}.
Here $\ITQFT_{3,4}(\mathcal{T})$ can be replaced by $2,3,4$- or $1,2,3,4$-theories alike.
In other words, a nonextended invertible TQFT uniquely extends to a $1,2,3,4$-TQFT with target $\mathcal{T}$.
The ambiguity of extending down to points then consists of a choice of an element in $\pi_0 (\Pic \mathcal{T})$ and an element of $\Z/6$.
This generalizes the second part of \cite[Theorem 7.6.3]{schommer2024invertible}.

\subsection{Other target categories and non-semisimple Crane--Yetter}

In this note, our main focus is target spectra with $\pi_1,\pi_2$ and $\pi_3$ all trivial.
Our arguments would apply to $\BrTens$ if it can be shown that $\pi_1 (\Pic \BrTens) = 0$.
If so, we would obtain the same classification result replacing $\mathcal{W}$ by $\pi_0(\Pic \Tens)$.
This is especially relevant to understanding Crane--Yetter for non-semisimple invertible braided tensor categories, which has recently attracted a lot of attention~\cite{kinnear2024nonsemisimple,costantino2023skein, brown2024skein}.

There are many other relevant target spectra, such as the universal target $\mathcal{U}_4$ as well as super versions of $\BrFus$ and $\BrTens$.
For such targets, the Picard $\infty$-groupoid has both $\pi_2$ and $\pi_3$ equal to $\Z/2$.
The universal property of $\Pic \mathcal{U}_4 = \pi_{\geq 0} \Sigma^4 I\C^\times$ ensures that there is exactly one invertible fully extended four-dimensional TQFT with partition function $Z_{(\lambda_1, \lambda_2)}$, and so only one fully extended Crane--Yetter with target $\mathcal{U}_4$.
We leave the study of other target spectra with nontrivial $\pi_1,\pi_2$ or $\pi_3$ to future work.

\bibliography{biblio.bib}{}
\bibliographystyle{plain}

\end{document}